\documentclass[11pt,a4paper,twoside]{article}
\usepackage[dvips]{graphics}
\usepackage{cite}
\usepackage{epsfig}
\usepackage{pennames}
\newcommand{\deriv}  {\ensuremath{\mathrm{d}}}

\begin{document}
%
\begin{titlepage}
\vspace*{-10mm}
\hbox to \textwidth{ \hsize=\textwidth
\hspace*{0pt\hfill} 
\vbox{ \hsize=58mm
{
\hbox{ MPI-PhE/2001-11 \hss}
\hbox{ June 15, 2001\hss } 
}
}
}

\bigskip\bigskip\bigskip
\begin{center}
{\huge\bf Measurement of the \\[1.5mm]
          longitudinal and transverse cross-section \\[1.5mm]
          in e$^{\mathbf{+}}$e$^{\mathbf{-}}$ annihilation
          at $\mathbf{\sqrt{s}=35}$-$\mathbf{44}$ GeV
}
\end{center}
\bigskip\bigskip
\begin{center}
{\Large  M.~Blumenstengel$^{(1)}$, O.~Biebel$^{(1)}$, 
         P.A.~Movilla~Fern\'andez$^{(1)}$, 
         P.~Pfeifenschneider$^{(1, a)}$,
         S.~Bethke$^{(1)}$, S.~Kluth$^{(1)}$
         and the JADE Collaboration$^{(2)}$
}
\end{center}
\bigskip
%
\begin{abstract}
\noindent 
An investigation of the polar angle distribution of charged
hadrons is presented using data taken by the JADE experiment 
at the PETRA \Pep\Pem\ collider at centre-of-mass energies 
of $35$ and $44$~GeV. From fits to the polar angle distribution 
the longitudinal, $\sigma_L$, and transverse, $\sigma_T$, 
cross-section relative to the total hadronic are determined 
at an average energy scale of $36.6$~GeV. 
The results are 
\begin{displaymath}
\frac{\sigma_L}{\sigma_{\mathrm{tot}}} = 0.067 \pm 0.013
\ \ ,
\mbox{\hspace*{1cm}}
\frac{\sigma_T}{\sigma_{\mathrm{tot}}} = 0.933 \mp 0.013
\end{displaymath}
where total errors are given and the results are exactly anti-correlated.
Using the next-to-leading order QCD prediction for the longitudinal 
cross-section, the value 
\begin{displaymath}
   \alpha_S(36.6\ {\mathrm{GeV}}) = 0.150 \pm 0.025
\end{displaymath}
of the strong coupling constant is obtained in agreement with the
world average value of $\alpha_S$ evolved to an energy scale of
$36.6$~GeV.
\end{abstract}

\vspace*{0pt\vfill}
\vfill
\begin{center}
\large Submitted to Phys.\ Lett.\ B
\end{center}
\bigskip

{
\small
\noindent
$^{(1)}$ 
\begin{minipage}[t]{155mm} 
Max-Planck-Institut f\"ur Physik, D-80805 M\"unchen, Germany \\
contact e-mail: biebel@mppmu.mpg.de 
\end{minipage}\\
$^{(2)}$ 
\begin{minipage}[t]{155mm} 
for a full list of members of the JADE Collaboration 
see Reference~\cite{bib-naroska}
\end{minipage}\\
$^{(a)}$  
\begin{minipage}[t]{155mm} 
now at SAP Deutschland AG \& Co.\ KG, Neurottstra{\ss}e 15, 69190 Walldorf, Germany
\end{minipage}
\hspace*{0pt\hfill}
}

\end{titlepage}
%
%
\newpage
\section{Introduction}
\label{sec-intro}
The energy and the momentum spectrum of a hadron $h$ produced in the annihilation 
process $\Pep\Pem \rightarrow \gamma, \PZz \rightarrow h + X$ is described by a 
fragmentation function 
${\cal F}^h(x) \equiv (1/\sigma_{\mathrm{tot}})\cdot (\deriv \sigma^h/\deriv x)$. 
Here $x$ is either the fractional momentum, $x_p \equiv 2 p /\sqrt{s}$, or 
the fractional energy, $x_E \equiv 2 E/\sqrt{s}$, carried by a hadron $h$, 
and $\sqrt{s}$ is the centre-of-mass energy of the annihilation process with 
total hadronic cross-section $\sigma_{\mathrm{tot}}$. In the case of 
unpolarized e$^\pm$ beams, and averaging over the polarization of the 
hadron $h$, the fragmentation function receives contributions from the 
transverse ($T$) and longitudinal ($L$) polarization states of the 
intermediate electroweak vector bosons, $\gamma$ and \PZz, and from their 
interference yielding an asymmetric contribution ($A$). The most general 
form of the differential cross-section for the inclusive single-particle 
production in \Pep\Pem\ annihilation is~\cite{bib-Nason-NPB421-473, 
bib-Altarelli-Mele-Nason}
\begin{equation}
\label{eqn-costheta-dependence}
    \frac{1}{\sigma_{\mathrm{tot}}} \cdot
             \frac{\deriv^2 \sigma^h}{\deriv x\ \deriv(\cos\theta)} =
    \frac{3}{8}\left(1+\cos^2\theta\right)\cdot {\cal F}_T^h(x)
  + \frac{3}{4}\left(\sin^2\theta\right)  \cdot {\cal F}_L^h(x)
  + \frac{3}{4}\left(\cos\theta\right)    \cdot {\cal F}_A^h(x)
 \ \ \ ,
\end{equation}
where $\theta$ is the polar angle between the direction of the incoming 
\Pem\ and the outgoing hadron $h$. At centre-of-mass energies much larger 
than the mass of the produced quark $q$, the longitudinal contribution 
is negligible~\cite{bib-BoehmHollik-CERN-89-08} due to the helicity structure 
at the quarks' production vertex. A sizeable contribution to the longitudinal 
fragmentation function, however, comes from gluon radiation from the $q\bar{q}$ 
system in the final state~\cite{bib-Nason-NPB421-473}. The asymmetric contribution 
is largest at energies below and above but very small at the Z peak. Even though it 
is $20$-$30$\% of the transverse cross-section \cite{bib-BoehmHollik-CERN-89-08} in 
the energy range of $35$ through $44$~GeV considered for this analysis, the 
required experimental distinction of quark and antiquark renders a measurement 
of the asymmetric contribution virtually impossible. It, therefore, was not 
considered for this analysis.

The fragmentation functions are related to the perturbatively calculable ratios 
of the longitudinal, $\sigma_L$, and transverse, $\sigma_T$, cross-sections 
to the total cross-section. Integrating Eq.~(\ref{eqn-costheta-dependence}) over
$\cos\theta$ and 
respecting energy conservation for the integral over $x$ yields 
\cite{bib-Nason-NPB421-473}
\begin{equation}
\label{eqn-sumrule}
    \frac{1}{2} \sum_h\int \deriv x\  x\cdot 
               \frac{1}{\sigma_{\mathrm{tot}}} \cdot
               \frac{\deriv \sigma^h}{\deriv x} = 
                         \frac{\sigma_{T}}{\sigma_{\mathrm{tot}}} + 
                         \frac{\sigma_{L}}{\sigma_{\mathrm{tot}}}     = 1
\ \ \ ,
\end{equation}
where
\begin{equation}
\label{eqn-sigma-def}
     \frac{\sigma_{T,L}}{\sigma_{\mathrm{tot}}} \equiv 
                         \frac{1}{2} \sum_h\int \deriv x\  x\cdot {\cal F}_{T,L}^h(x)
\ \ \ .
\end{equation}
The contribution of gluon radiation to $\sigma_{T,L}/\sigma_{\mathrm{tot}}$
has been calculated in second order of $\alpha_S$~\cite{bib-Rijken-PLB386-422}. 
At first order the QCD correction in the total cross-section 
contributes only to the longitudinal part \cite{bib-Nason-NPB421-473}:
\begin{equation}
\label{eqn-sigma_tot}
  \sigma_{\mathrm{tot}} = \sigma_T + \sigma_L 
                        = \left(                     \sigma_0 + {\cal O}(\alpha_S^2)\right) +
                          \left(\frac{\alpha_S}{\pi} \sigma_0 + {\cal O}(\alpha_S^2)\right) 
        \ \ \ ,
\end{equation}
where $\sigma_0$ is the Born level cross-section.
This allows tests of QCD and determinations of $\alpha_S$ from measurements
of $\sigma_L/\sigma_{\mathrm{tot}}$ and $\sigma_T/\sigma_{\mathrm{tot}}$.

The longitudinal and transverse fragmentation functions were already investigated 
by the SLAC/LBL magnetic detector collaboration at the SPEAR collider at 
$7.4$~GeV~\cite{bib-PRL35-1320}, by the TASSO collaboration at the PETRA 
collider at $14$, $22$, and $34$~GeV~\cite{bib-PLB114-65}, and by the OPAL, 
ALEPH, and DELPHI collaborations at the LEP collider at 
$\sqrt{s}\approx m_\PZz$~\cite{bib-OPAL-ZPC68-203, bib-ALEPH-PLB357-487, 
bib-DELPHI-EPJC6-19}. 
Ratios $\sigma_L/\sigma_{\mathrm{tot}}$ to derive $\alpha_S(m_\PZz)$ were 
only determined by the OPAL and DELPHI collaborations \cite{bib-OPAL-ZPC68-203, 
bib-DELPHI-EPJC6-19}.

Due to the sum rule Eq.~(\ref{eqn-sumrule}) all details of the unknown 
fragmentation functions disappear from the longitudinal and transverse 
cross-sections up to corrections suppressed by some power of $1/\sqrt{s}$. 
A measurement of these cross-sections at centre-of-mass energies different 
from $\sqrt{s} \approx m_\PZz$ is, therefore, indispensable to experimentally 
investigate the question whether power corrections to Eq.~(\ref{eqn-sigma_tot})
are required and whether these are of the form $1/\sqrt{s}$ or 
$1/s$~\cite{bib-sigma_L-powcorr}. 

The analysis presented in the following determines 
$\sigma_L/\sigma_{\mathrm{tot}}$, $\sigma_T/\sigma_{\mathrm{tot}}$,
and $\alpha_S$ at an average energy of $\langle\sqrt{s}\rangle = 36.6$~GeV.
It uses data measured with the JADE detector~\cite{bib-naroska,
bib-JADEdet} at the PETRA collider to be introduced in Section~\ref{sec-detector}.
The measurement of the $\cos\theta$ distribution and the investigation
of the experimental systematics are detailed in Section~\ref{sec-costheta}.
Section~\ref{sec-observables} presents the results from the fits 
to the measured $\cos\theta$ distribution. Power corrections to 
Eq.~(\ref{eqn-sigma_tot}) are considered in Section~\ref{sec-powercorrection}. 
Our final results for $\sigma_L/\sigma_{\mathrm{tot}}$, 
$\sigma_T/\sigma_{\mathrm{tot}}$, and $\alpha_S$ are summarized in 
Section~\ref{sec-summary}.

\section{Detector and data samples}
\label{sec-detector}
\begin{table}
\begin{center}
\renewcommand{\arraystretch}{1.5}
\begin{tabular}{|p{56mm}||l|}                                                             
\hline
description                & cut                                                        \\
\hline\hline
minimum track momentum        
                           & $p > 0.1$~GeV                                              \\
tracks coming out of a cylinder 
(\O~$3\,$cm $\times 7\,$cm) 
around the \Pep\Pem\ vertex   
                           & $n_{\mathrm{ch}}^{\mathrm{vertex}} \ge 4$                  \\
tracks having $\ge 24$ points 
and $p_t>500$~MeV             
                           & $n_{\mathrm{ch}} \ge 3$                                    \\
visible energy                
                           & $E_{\mathrm{vis}} = \sum_i E_i> \sqrt{s}/2$                \\
longitudinal momentum balance 
                           & $p_{\mathrm{bal}} = |\sum p^z_i/E_{\mathrm{vis}}| < 0.4$   \\
axial vertex position         
                           & $|z_{\mathrm{VTX}}| < 150$~mm                              \\
polar angle of thrust axis    
                           & $|\cos\theta_{T}| < 0.8$                                   \\
total missing momentum        
                           & $p_{\mathrm{miss}} = |\sum \vec{p}_i| < 0.3\cdot \sqrt{s}$ \\
\hline
\end{tabular}
\end{center}
\caption{\label{tab-selection-cuts} The main cuts are listed for the selection 
         of multihadronic events which were varied
         to assess systematic uncertainties (see text). 
         $E_i$ and $\vec{p}_i$ are energy and 3-momentum of tracks and 
         clusters.
        }
\end{table}
\begin{table}
\begin{center}
\begin{tabular}{|r|c||c|c|}
\hline
year & $\sqrt{s}\ [\mathrm{GeV}]$ & data & MC \\ 
\hline\hline
1979-1985 &  $34$-$36$     &              $13\thinspace 013$ & $19\thinspace 814$            \\
     1986 &  $34$-$36$     &              $20\thinspace 926$ & $25\thinspace 123$            \\
  1984/85 &  $43$-$45$     &                          $4504$ & $14\thinspace 497$            \\ 
\hline
\end{tabular}
\end{center}
\caption{\label{tab-eventnumbers}
         Number of selected multihadronic events in data and Monte Carlo detector
         simulation (MC).
}
\end{table}
The investigation presented in this paper is a re-analysis of data
recorded by the JADE detector at the PETRA electron-positron collider.
The JADE detector is described in detail elsewhere~\cite{bib-naroska,
bib-JADEdet}. The main component of the detector used for this study is the 
central jet chamber which measured the tracks of charged particles with 
$8$ up to $48$ points in about 97\% of the full solid angle. The relative 
resolution of the transverse track momentum was 
   $\sigma(p_t)/p_t = \sqrt{0.04^2+(0.018\cdot p_t [\mathrm{GeV}/c])^2}$.
The spatial resolution in the $r$-$\varphi$ plane\footnote{JADE used a 
cylinder coordinate system with the $z$ axis along the \Pem\ beam direction, 
the radius $r$ is the distance from the $z$ axis, the azimuthal angle $\varphi$
is measured from the horizontal plane, and the polar angle $\theta$ is
measured with respect to the $z$ axis.} was $180\ \mu$m before 1986 and
$110\ \mu$m for the data measured in 1986 due to the installation of a
digital readout system. The resolution along the $z$ axis was $1.6$~cm
which degraded when the digital readout came into operation.

The data used for this study were recorded between 1979 and 1986 at 
centre-of-mass energies of $\sqrt{s} = 34$-$36$~GeV and $\sqrt{s} = 43$-$45$~GeV. 
Multihadronic events were selected according to the criteria described in
\cite{bib-Fernandez-EPJC1-461}. Applying the selection cuts listed in 
Tab.~\ref{tab-selection-cuts} yielded the number of events in data and Monte Carlo 
simulation (MC) listed in Tab.~\ref{tab-eventnumbers}.
As in our previous publication \cite{bib-Fernandez-EPJC1-461} we used the
JADE collaboration's original Monte Carlo samples of multihadronic events 
from the JETSET program version 6.3~\cite{bib-JETSET} including a detailed 
simulation of the JADE detector which were available for these energies. 

Since the data at $\sqrt{s} = 34$-$36$~GeV were recorded with two 
different configurations of the JADE detector (see~\cite{bib-naroska}),
the distributions of the polar angle obtained from these data sets were 
corrected separately for detector effects. After noticing a good 
agreement of the corrected distributions they were combined for the fits.

\begin{figure}
\includegraphics[width=0.5\textwidth]{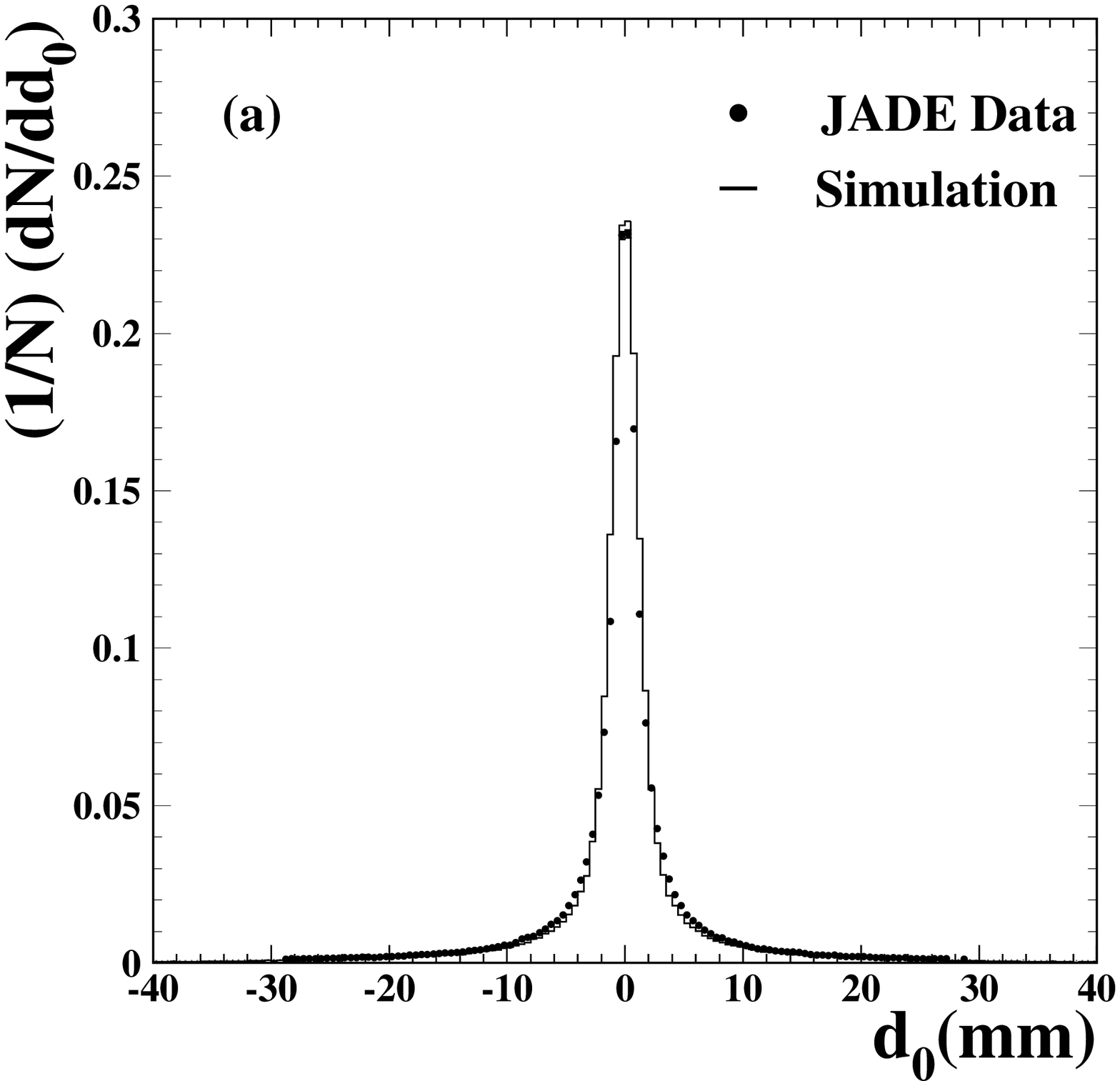}
\includegraphics[width=0.5\textwidth]{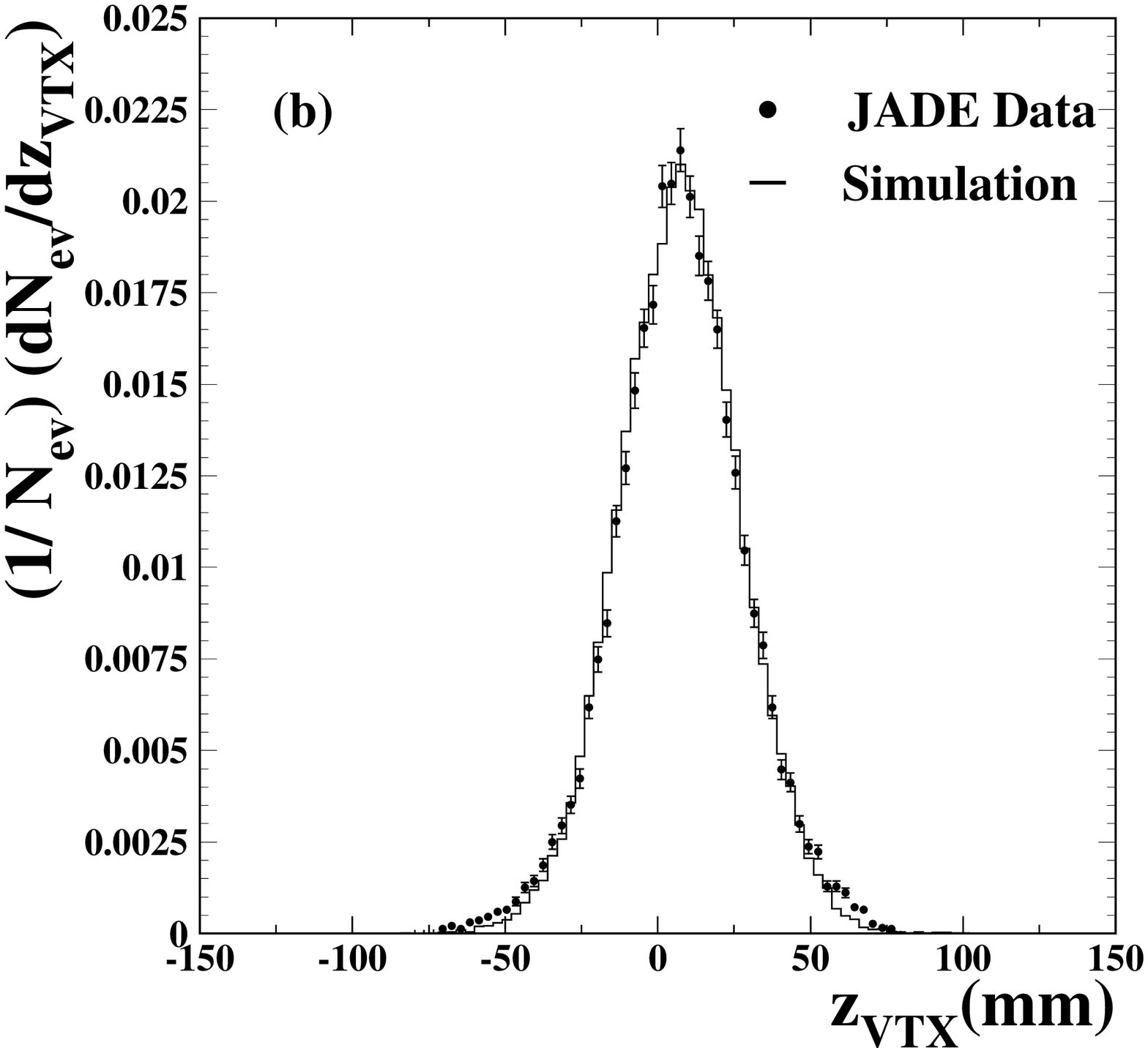}
\caption{\label{fig-DOCA-zVTX}
         The distributions of (a) the tracks' radial distance of closest 
         approach to the vertex, $d_0$, and (b) the vertex' axial position 
         are shown for data (points) and simulation (histogram) after 
         shifting and smearing (see text).
}
\end{figure}
The simulated data had to be adapted to the experimental position of the 
\Pep\Pem\ collision point (I.P.) and the resolutions of the measured 
$z$ vertex position, $z_{\mathrm{VTX}}$, and of the minimum radial distance 
of a track to the I.P., $d_0$.
Fig.~\ref{fig-DOCA-zVTX} shows the comparison of the smeared $d_0$ and
$z_{\mathrm{VTX}}$ distributions in the simulation and in the data. In 
radial direction only a gaussian smearing of $0.8$~mm was required while 
for the $z$ vertex position a smearing of $17.5$~mm and a shift of $7.4$~mm 
were needed.

\section{Measurement of $\mathbf{\cos\theta}$ distribution}
\label{sec-costheta}
\begin{figure}
\includegraphics[width=1.00\textwidth]{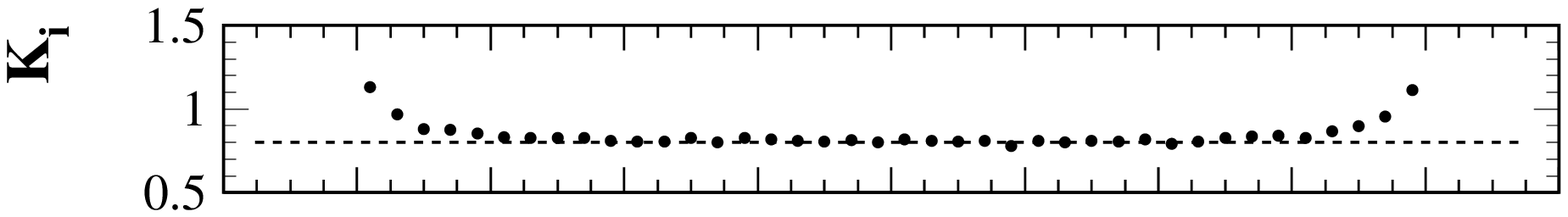}
\includegraphics[width=1.00\textwidth]{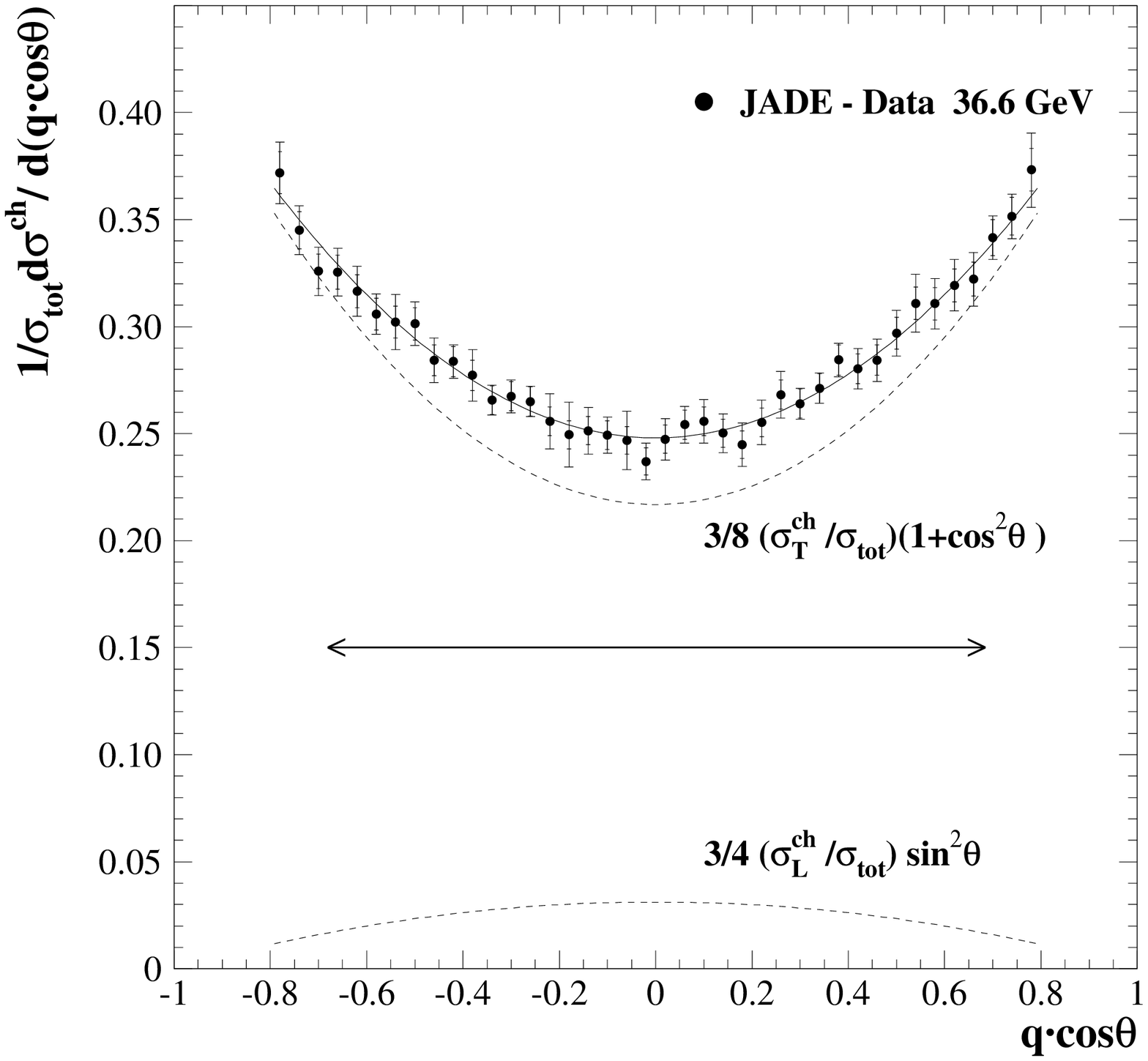}
\caption{\label{fig-costheta}
         The distribution of $q\cdot\cos\theta$ is shown after correction for 
         detector effects using the factors $K_i$ presented in the upper part 
         of the figure. The inner error bars are the statistical uncertainties 
         due to data and limited detector simulation statistics, and the outer bars 
         are the total errors. The range considered for the fit is indicated by the 
         arrow.
}
\end{figure}
To determine the longitudinal and transverse cross-sections the distribution 
of $\cos \theta$ of all tracks of charged particles 
was measured. 
The small correlation of the quark's and the particle's charge indispensible
for a determination of the asymmetric cross-section is maintained by multiplying
$\cos\theta$ of each particle by the sign $q$ of the particle's charge. Since
the experimental sensitivity on the asymmetric contribution is marginal, the 
fits considered only the
longitudinal and transverse contribution to the cross-section which are 
insensitive to the sign of $q\cdot\cos\theta$.

The effects of limited acceptance and resolution of the detector were corrected
using a bin-by-bin correction method. 
\begin{table}
\renewcommand{\arraystretch}{1.4}
\begin{center}
\begin{tabular}{|r@{~-~}r|r@{~$\pm$~}r@{~$\pm$~}r|c|r@{~-~}r|r@{~$\pm$~}r@{~$\pm$~}r|}
\cline{1-5}
\cline{7-11}
\multicolumn{2}{|c|}
{$q\cdot\cos\theta$ range} & 
      \multicolumn{3}{|c|}{$\frac{1}{\sigma_{\mathrm{tot}}}\frac{\deriv \sigma^{\mathrm{ch}}}{\deriv (q\cdot\cos\theta)}$} &
                        \hspace*{1cm} &
\multicolumn{2}{|c|}
{$q\cdot\cos\theta$ range} & 
      \multicolumn{3}{|c|}{$\frac{1}{\sigma_{\mathrm{tot}}}\frac{\deriv \sigma^{\mathrm{ch}}}{\deriv (q\cdot\cos\theta)}$} \\
\cline{1-5}\cline{1-5}
\cline{7-11}\cline{7-11}
%
%
$-0.80$ & $-0.76$ & $0.372$ & $0.010$ & $0.011$  && ~~
                                                    $ 0.00$ & $ 0.04$ & $0.247$ & $0.007$ & $0.007$  \\ 
$-0.76$ & $-0.72$ & $0.345$ & $0.009$ & $0.007$  && $ 0.04$ & $ 0.08$ & $0.254$ & $0.007$ & $0.005$  \\ 
$-0.72$ & $-0.68$ & $0.326$ & $0.008$ & $0.008$  && $ 0.08$ & $ 0.12$ & $0.256$ & $0.007$ & $0.008$  \\ 
$-0.68$ & $-0.64$ & $0.326$ & $0.008$ & $0.008$  && $ 0.12$ & $ 0.16$ & $0.250$ & $0.007$ & $0.006$  \\ 
$-0.64$ & $-0.60$ & $0.316$ & $0.008$ & $0.009$  && $ 0.16$ & $ 0.20$ & $0.245$ & $0.006$ & $0.008$  \\ 
$-0.60$ & $-0.56$ & $0.306$ & $0.007$ & $0.006$  && $ 0.20$ & $ 0.24$ & $0.255$ & $0.007$ & $0.008$  \\ 
$-0.56$ & $-0.52$ & $0.302$ & $0.007$ & $0.011$  && $ 0.24$ & $ 0.28$ & $0.268$ & $0.007$ & $0.009$  \\ 
$-0.52$ & $-0.48$ & $0.301$ & $0.007$ & $0.007$  && $ 0.28$ & $ 0.32$ & $0.264$ & $0.007$ & $0.003$  \\ 
$-0.48$ & $-0.44$ & $0.284$ & $0.007$ & $0.008$  && $ 0.32$ & $ 0.36$ & $0.271$ & $0.007$ & $0.002$  \\ 
$-0.44$ & $-0.40$ & $0.284$ & $0.007$ & $0.003$  && $ 0.36$ & $ 0.40$ & $0.284$ & $0.007$ & $0.003$  \\ 
$-0.40$ & $-0.36$ & $0.277$ & $0.007$ & $0.010$  && $ 0.40$ & $ 0.44$ & $0.280$ & $0.007$ & $0.006$  \\ 
$-0.36$ & $-0.32$ & $0.266$ & $0.007$ & $0.002$  && $ 0.44$ & $ 0.48$ & $0.284$ & $0.007$ & $0.007$  \\ 
$-0.32$ & $-0.28$ & $0.267$ & $0.007$ & $0.004$  && $ 0.48$ & $ 0.52$ & $0.297$ & $0.007$ & $0.008$  \\ 
$-0.28$ & $-0.24$ & $0.265$ & $0.007$ & $0.002$  && $ 0.52$ & $ 0.56$ & $0.311$ & $0.008$ & $0.011$  \\ 
$-0.24$ & $-0.20$ & $0.256$ & $0.007$ & $0.011$  && $ 0.56$ & $ 0.60$ & $0.311$ & $0.008$ & $0.009$  \\ 
$-0.20$ & $-0.16$ & $0.250$ & $0.007$ & $0.014$  && $ 0.60$ & $ 0.64$ & $0.319$ & $0.008$ & $0.009$  \\ 
$-0.16$ & $-0.12$ & $0.251$ & $0.007$ & $0.009$  && $ 0.64$ & $ 0.68$ & $0.322$ & $0.008$ & $0.010$  \\ 
$-0.12$ & $-0.08$ & $0.249$ & $0.007$ & $0.005$  && $ 0.68$ & $ 0.72$ & $0.342$ & $0.009$ & $0.006$  \\ 
$-0.08$ & $-0.04$ & $0.247$ & $0.007$ & $0.012$  && $ 0.72$ & $ 0.76$ & $0.352$ & $0.009$ & $0.006$  \\ 
$-0.04$ & $ 0.00$ & $0.237$ & $0.006$ & $0.006$  && $ 0.76$ & $ 0.80$ & $0.373$ & $0.010$ & $0.014$  \\ 
\cline{1-5}
\cline{7-11}
\end{tabular}
\end{center}
\caption{\label{tab-costheta}
         The differential $q\cdot\cos\theta$ distribution data are listed 
         for charged particles with statistical and systematic uncertainties. 
         The statistical uncertainties include the uncertainties due to the
         limited statistics of the detector simulation.
}
\end{table}
The correction was obtained from the 
detailed detector simulation as the binwise ratio of the $q\cdot\cos\theta$ 
distribution at the hadron level and the corresponding distribution at the 
detector level. Here, hadron level means all charged particles having lifetimes 
greater than $300$~ps generated by the Monte Carlo event generator, and the 
detector level comprises all charged particles that were observed after passing 
the simulated events through the detector simulation and reconstruction programs. 
Effects due to the neutral particles were not corrected using the simulation but 
were obtained from the measured data as will be detailed in Section~\ref{sec-observables}. 
Fig.~\ref{fig-costheta} shows the measured distribution of $q\cdot\cos\theta$ 
after application of the binwise correction factors 
which are shown in the upper part of the figure. 
In the central part of the detector the distribution of the correction factors 
is flat at about $0.8$ and increases towards the acceptance boundaries at large
$|\cos\theta|$. The values are below unity due to normalizing all distributions
to the mean charged multiplicity, in particular those at detector level where 
the acceptance is reduced by the cut on the polar angle of the thrust axis.

All data sets measured at centre-of-mass energies of about $35$ and $44$~GeV 
and corrected for detector effects were combined, weighted with the 
respective integrated luminosities. The measured values and the statistical 
and systematic uncertainties are listed in Tab.~\ref{tab-costheta}.
The corrected distribution is limited to the range of $|\cos\theta|<0.8$
due to the acceptance limit implied by the cut on the polar angle of the 
thrust axis.

As an additional cross-check we determined the mean charged multiplicity
from the corrected distribution at $35$~GeV, yielding
$\langle n_{\mathrm{ch}}\rangle = 14.23\pm 0.04$ where the error is 
statistical only. This is in agreement within the total error 
of the published result 
$\langle n_{\mathrm{ch}}\rangle=13.6 \pm 0.3 ({\mathrm{stat.}}) \pm 0.6 ({\mathrm{syst.}})$
\cite{bib-zpc20-187} at this energy
where tracks were counted by visually inspecting the events.

To assess the systematic uncertainties due to imperfections of the detector simulation,
and due to the contributions from background processes, the main selection cuts listed 
in Tab.~\ref{tab-selection-cuts} were varied. The measurement of the $q\cdot\cos\theta$ 
distribution was repeated for each variation and any deviation from the distribution 
obtained using the standard selection cuts was considered a systematic uncertainty. 
Tab.~\ref{tab-alpha_S-errors} summarises all investigated variations of the selection 
cuts.

Deficiencies of the description of the data by the simulation were considered as a 
source of systematic uncertainty. The largest deviation to be considered for this 
investigation is due to the choice of the fragmentation function which yielded
more high energetic particles than observed in the data. Its contribution to the 
systematic error of the $q\cdot\cos\theta$ distribution was obtained from reweighting 
the simulation to match the fragmentation function of the data prior to determining 
the correction factors.

Finally, additional selection cuts on $d_0$ were applied to reject tracks which stem 
from decays of long-lived particles or from interactions with the detector material.  
A similar cut on the axial distance of a track to the reconstructed vertex position,
$z_0$, could not be applied since the relevant information is not available in the 
preprocessed data files~\cite{bib-Fernandez-EPJC1-461} we used. Instead we varied the 
selection cut on the event-by-event vertex position, $z_{\mathrm{VTX}}$, about its 
average derived from all events, $\langle z_{\mathrm{VTX}}\rangle$.  
The systematic uncertainty was found from applying tighter cuts on $d_0$ and
$z_{\mathrm{VTX}}$ and repeating the measurement. The cuts, $d_0<3$~mm or $19$~mm, 
and $|z_{\mathrm{VTX}}-\langle z_{\mathrm{VTX}}\rangle| <29$~mm or $39$~mm, were 
derived from one and a half and twice the gaussian width of the corresponding 
distribution measured in the data. 
The looser cut on $d_0$, however, was found from fitting an exponential function
to the $d_0$ distribution at large values of $d_0$. 
The fit was extended to the largest range describable with this exponential 
function. The lower end of this fit range was chosen for the looser cut on $d_0$.
This considers that large positive
values of $d_0$ are dominantly due to tracks from decays of long-lived particles 
such as K$^0_S$ and $\Lambda$.

\section{Determination of $\mathbf{\sigma_L/\sigma_{\mathrm{\bf tot}}}$, 
                          $\mathbf{\sigma_T/\sigma_{\mathrm{\bf tot}}}$,
                  and $\mathbf{\alpha_S}$
}
\label{sec-observables}
From the $q\cdot\cos\theta$ distribution $\sigma_L/\sigma_{\mathrm{tot}}$
and $\sigma_T/\sigma_{\mathrm{tot}}$ can be determined after neutral 
particles not included in the $q\cdot\cos\theta$ distribution are
taken into account. Studies using the JETSET Monte Carlo 
generator~\cite{bib-JETSET} at $\sqrt{s}=35$~GeV with the parameters
quoted in~\cite{bib-Fernandez-EPJC1-461,bib-JADE-JETSET-tune} yielded 
\begin{eqnarray}
\label{eqn-JETSET-eta^ch}
 \left(\eta_L^{\mathrm{ch}}\right)^{\mathrm{had,MC}} \equiv
 \left(\frac{\sigma_L^{\mathrm{ch}}}{\sigma_L}\right)^{\mathrm{had,MC}} & = & 0.608 \pm 0.004, 
        \nonumber \\
 \left(\eta_T^{\mathrm{ch}}\right)^{\mathrm{had,MC}} \equiv
 \left(\frac{\sigma_T^{\mathrm{ch}}}{\sigma_T}\right)^{\mathrm{had,MC}} & = & 0.6179 \pm 0.0004
\end{eqnarray}
for the ratio of cross-sections obtained from charged and from charged plus 
neutral particles. Nearly identical ratios were found 
in 
\cite{bib-OPAL-ZPC68-203} using JETSET at 
$\sqrt{s}=91.2$~GeV with a different parameter set~\cite{bib-OPAL-ZPC69-543}. 
Thus, independently of the centre-of-mass energy, the 
correction of the longitudinal cross-section for neutral particles is expected 
to be about $1.6\%$ larger than the same correction for the transverse cross-section. 
Since the correction depends on the details of the hadronization model,
and since the absolute difference of the corrections in Eq.~(\ref{eqn-JETSET-eta^ch})
is less than the statistical uncertainty of our measurement, the difference 
was neglected for the determination of the longitudinal and transverse cross-section 
but was considered for the systematic uncertainties. 

Assuming that at the hadron level the correction for
neutral particles is identical for the longitudinal and transverse cross-sections, 
using Eq.~(\ref{eqn-sigma-def}) the differential cross-section for charged particles 
given by Eq.~(\ref{eqn-costheta-dependence}) can be written as
\begin{equation}
\label{eqn-sigmaL-fit}
    \frac{1}{\sigma_{\mathrm{tot}}} \cdot
             \frac{\deriv \sigma^{\mathrm{ch}}}{\deriv (q\cdot\cos\theta)} =
    \frac{3}{8}\eta^{\mathrm{ch}}
               \left[\frac{\sigma_L}{\sigma_{\mathrm{tot}}}\left(1-3\cos^2\theta\right)
                                                         + \left(1+ \cos^2\theta\right)
               \right]
\ \ \ .
\end{equation}
The unknown parameters to be determined from a fit to the data are 
$\eta^{\mathrm{ch}}$, which is the correction factor for the
total cross-section accounting for the neutral particles, and 
$\sigma_L/\sigma_{\mathrm{tot}}$. From the relation known in 
${\cal O}(\alpha_S^2)$ \cite{bib-Rijken-PLB386-422}
\begin{equation}
\label{eqn-QCD-prediction}
\left( \frac{\sigma_L}{\sigma_{\mathrm{tot}}}\right)_{\mathrm{PT}} =
                     \frac{\alpha_S}{\pi} + 8.444\left(\frac{\alpha_S}{\pi}\right)^2
\end{equation}
a formula similar to Eq.~(\ref{eqn-sigmaL-fit}) can be derived which 
allows for a direct determination of the strong coupling constant $\alpha_S$.

The largest sensitivity to the longitudinal cross-section comes from 
the central region, $|\cos\theta| \approx 0$, and the forward regions, 
$|\cos\theta| \rightarrow 1$. Since measurements in the forward region 
are affected by the limited detector acceptance, the central part of 
the detector, $|\cos\theta|<0.68$, was chosen for the range of the fit. 
For this fit range a good fit was obtained with 
$\chi^2/{\mathrm{d.o.f.}}\approx 0.49$.\footnote{Using the statistical
errors of the data only gave $\chi^2/{\mathrm{d.o.f.}}\approx 0.89$}
The two fits yielded 
\begin{equation}
\label{eqn-fit-results}
      \frac{\sigma_L}{\sigma_{\mathrm{tot}}}   =   0.067 \pm 0.011 
\mbox{\ \ \ and\ \ \ }                                              \nonumber \\
      \alpha_S(36.6\ {\mathrm{GeV}})           =   0.150 \pm 0.020            
\end{equation}
at the luminosity weighted average centre-of-mass energy of $36.6$~GeV
where the errors are from the fit, and where $\eta^{\mathrm{ch}}=0.6196 \pm 0.0043$ 
was obtained in both fits. 
The fitted value of $\eta^{\mathrm{ch}}$ agrees within errors with the
expectation of the JETSET Monte Carlo generator in 
Eq.~(\ref{eqn-JETSET-eta^ch}). A significant correlation
of $\eta^{\mathrm{ch}}$ with $\alpha_S$ and $\sigma_L/\sigma_{\mathrm{tot}}$ 
of $-77\%$ is present. The correlation increases 
when smaller fit ranges are chosen. This signals a reduced ability of 
the fit to distinguish the longitudinal contribution to the cross-section 
from a simple change of the normalization.
The result for the transverse cross-section can be derived from
Eq.~(\ref{eqn-fit-results}) with Eq.~(\ref{eqn-sigma_tot}) and
is therefore exactly anti-correlated to the longitudinal 
cross-section.

Besides the errors propagated from the measured $q\cdot\cos\theta$ 
distribution the fits were repeated for every systematic variation 
of the measurement. Deviations with respect to the standard fit 
results were taken as systematic uncertainties. Several other fit 
ranges, $|\cos\theta|<0.52 \ldots 0.80$ were considered. The largest 
up- and downward excursion from the standard result was assigned as 
the uncertainty due to the choice of the fit range. Due to the 
correlation between the two fit parameters, the value of
$\eta^{\mathrm{ch}}$ was kept fixed at $0.6196$
for the variation of the fit ranges. Otherwise the reduced discrimination 
power between 
$\eta^{\mathrm{ch}}$ and $\alpha_S$ ($\sigma_L/\sigma_{\mathrm{tot}}$)
for small fit ranges biases the estimation of this uncertainty. 
The $1.6\%$ difference in the correction of the
longitudinal and transverse cross-sections for the contribution of 
neutral particles was examined by introducing $\eta^{\mathrm{ch}}_L$ 
and $\eta^{\mathrm{ch}}_T$ obeying the relation $\eta^{\mathrm{ch}}_T = 1.016 
\eta^{\mathrm{ch}}_L$ in the fit formula Eq.~(\ref{eqn-sigmaL-fit}).
This yielded a negligible contribution to the overall systematic 
uncertainties. For the $\alpha_S$ determination an additional 
uncertainty arises from the choice of the renormalization scale, 
which was varied from $\mu=\sqrt{s}$ by a factor 
$x_\mu\equiv \mu/\sqrt{s}=0.5$ or $2$.
The individual positive and negative systematic error contributions 
were added in quadrature and symmetrized for the final result.

The results with all alloted errors are
\begin{eqnarray}
\label{eqn-full-results}
      \frac{\sigma_L}{\sigma_{\mathrm{tot}}} & = & 0.067  \pm 0.011  ({\mathrm{stat.}}) \pm 0.007  ({\mathrm{syst.}}) \nonumber \\
      \alpha_S(36.6\ {\mathrm{GeV}})         & = & 0.150  \pm 0.020  ({\mathrm{stat.}}) \pm 0.013  ({\mathrm{syst.}}) 
                                                                                        \pm 0.008  ({\mathrm{scale}}) 
\ \ \ ,
\end{eqnarray}
where the third error on $\alpha_S$ is due to the variation of 
the renormalization scale. Tab.~\ref{tab-alpha_S-errors} shows 
the individual error contributions to the results which are 
dominated by the statistical error.
\begin{table}
\renewcommand{\arraystretch}{1.4}
\begin{center}
\begin{tabular}{|l||r|r||r|}
\cline{2-4}
 \multicolumn{1}{c|}{}
                 & $\sigma_L/\sigma_{\mathrm{tot}}$ 
                                   & $\sigma_T/\sigma_{\mathrm{tot}}$ 
                                                     & $\alpha_S(36.6\ {\mathrm{GeV}})$       \\
\cline{2-4}\hline
fit result       &         $0.067$ &         $0.933$ &         $0.150$  \\
\hline\hline
data statistics
                 &     $\pm 0.009$ &     $\mp 0.009$ &     $\pm 0.016$  \\
MC statistics
                 &     $\pm 0.007$ &     $\mp 0.007$ &     $\pm 0.012$  \\
\hline
total stat.\ error     
                 &     $\pm 0.011$ &     $\mp 0.011$ &     $\pm 0.020$  \\
\hline\hline
$p > 0.2$~GeV 
                 &     $   <0.001$ &     $   <0.001$ &     $   <0.001$  \\
$n_{\mathrm{ch}}^{\mathrm{vertex}} > 7$ 
                 &     $   <0.001$ &     $   <0.001$ &     $   +0.001$  \\
$n_{\mathrm{ch}} > 7$ 
                 &     $   +0.006$ &     $   -0.006$ &     $   +0.011$  \\
$E_{\mathrm{vis}} > (0.95 \ldots 1.05)\cdot\sqrt{s}/2$
                 &     $\pm 0.001$ &     $\mp 0.001$ &     $\pm 0.001$  \\
$p_{\mathrm{bal}} < (0.3 \ldots \infty)$
                 &     $\pm 0.002$ &     $\mp 0.002$ &     $\pm 0.004$  \\
$|z_{\mathrm{VTX}}-\langle z_{\mathrm{VTX}}\rangle| < (29\ldots 39)$~mm
                 &     $   +0.002$ &     $   -0.002$ &     $   +0.004$  \\
$|\cos\theta_T| < (0.7 \ldots 0.9)$   
                 &     $^{\textstyle +0.001}_{\textstyle -0.006}$ 
                                   &     $^{\textstyle -0.001}_{\textstyle +0.006}$ 
                                                     &     $^{\textstyle +0.003}_{\textstyle -0.010}$  \\
$p_{\mathrm{miss}}< (0.25 \ldots \infty)\cdot\sqrt{s}$
                 &     $\pm 0.001$ &     $\mp 0.001$ &     $\pm 0.002$  \\
$d_0 <  (3 \ldots 19)$~mm
                 &     $   -0.001$ &     $   +0.001$ &     $   -0.003$  \\
\cline{1-1}
$\eta_T^{\mathrm{ch}} = 1.016\cdot\eta_L^{\mathrm{ch}}$
                 &     $   <0.001$ &     $   <0.001$ &     $   <0.001$  \\
reweighting of fragmentation fct.
                 &     $   +0.002$ &     $   -0.002$ &     $   +0.005$  \\
\cline{1-1}
fit range $|\cos\theta| < 0.52 \ldots 0.8$        
                 &     $\pm 0.001$ &     $\pm 0.001$ &     $\pm 0.002$  \\
\hline
total syst.\ error
                 &     $\pm 0.007$
                                   &     $\mp 0.007$
                                                      &     $\pm 0.013$ \\
\hline\hline
$x_{\mu}=0.5\ldots 2$
                 &             --- &              --- &     $\pm 0.008$ \\
\hline\hline
total error
                 &     $\pm 0.013$
                                   &     $\mp 0.013$
                                                      &     $\pm 0.025$ \\
\hline
\end{tabular}
\end{center}
\caption{\label{tab-alpha_S-errors} 
         Error contributions for the determinations of $\sigma_L/\sigma_{\mathrm{tot}}$, 
         $\sigma_T/\sigma_{\mathrm{tot}}$, which are exactly anti-correlated, and of 
         $\alpha_S(36.6\ {\mathrm{GeV}})$. 
}
\end{table}
In this table, the contributions of the statistical uncertainties due to
data and Monte Carlo simulation are given separately to illustrate the
possible gain from a larger sample of simulated events on the total errors.

\section{Power correction}
\label{sec-powercorrection}
\begin{figure}
\includegraphics[width=1.00\textwidth]{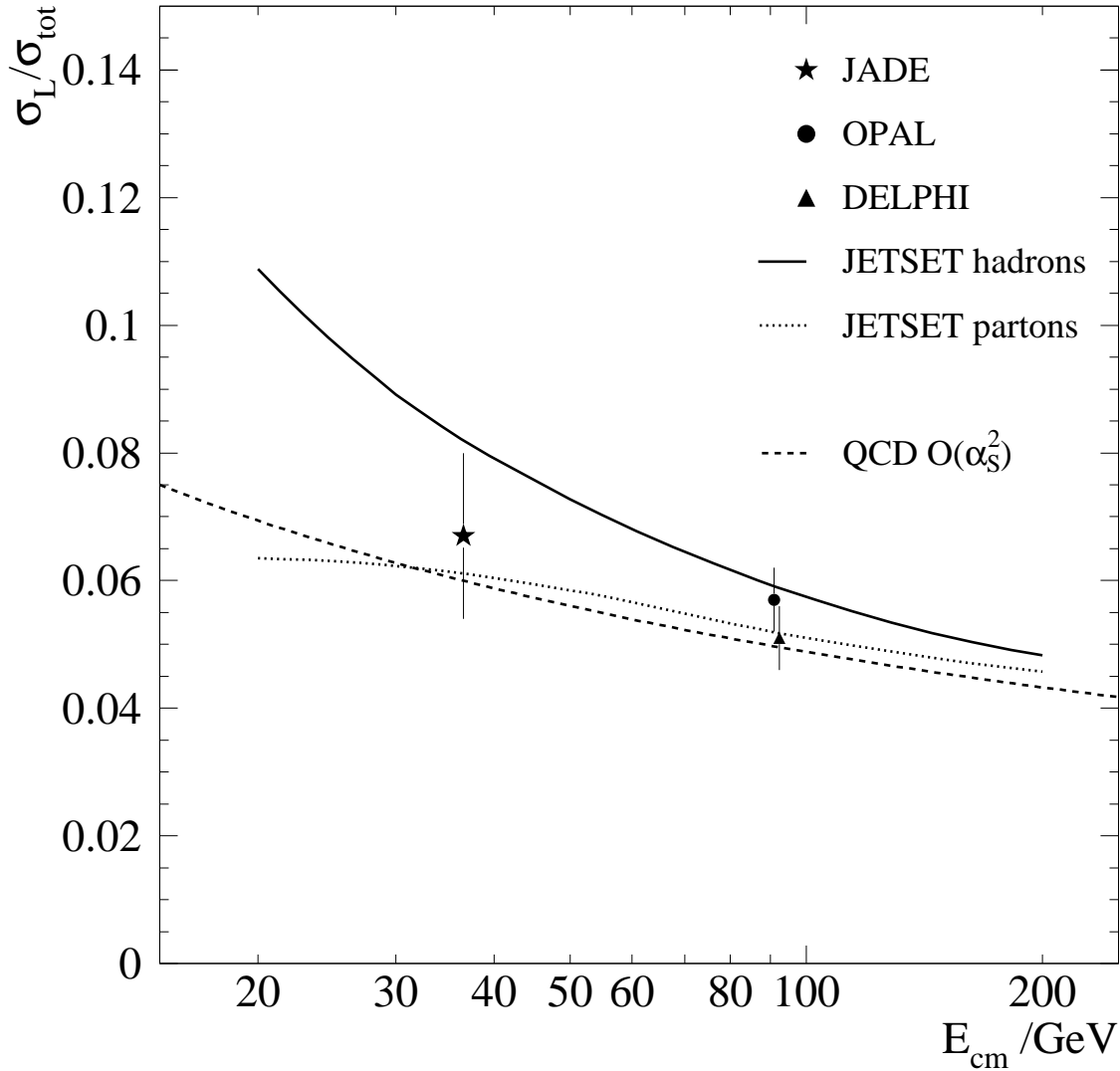}
\caption{\label{fig-sigmaL-vs-Q}
         Longitudinal cross-section relative to the total cross-section
         is presented versus centre-of-mass energy. 
         The results obtained 
         at the Z peak~\cite{bib-OPAL-ZPC68-203, 
         bib-DELPHI-EPJC6-19} are also shown. 
         The dotted and solid lines show, respectively, the 
         JETSET\protect\footnotemark~\cite{bib-JETSET}
         expectation for partons and stable particles. 
         The dashed line is the second 
         order QCD prediction, Eq.~(\ref{eqn-QCD-prediction}), using
         $\alpha_S(m_{\mathrm{Z}})=0.1184 \pm 0.0031$
         \cite{bib-Bethke-0004021}.
}
\end{figure}

Fig.~\ref{fig-sigmaL-vs-Q} shows our result and those obtained at the Z 
peak~\cite{bib-OPAL-ZPC68-203, bib-DELPHI-EPJC6-19} for the ratio of 
the longitudinal and total cross-section versus the centre-of-mass 
energy. Neither the measurement of the SLAC/LBL magnetic detector \cite{bib-PRL35-1320}
nor that of the TASSO collaboration \cite{bib-PLB114-65} were considered 
for the study of power corrections. The results published by the SLAC/LBL 
collaboration, from which $\sigma_L/\sigma_{\mathrm{tot}}=0.10\pm 0.02$(stat.) 
at $\sqrt{s}=7.4$~GeV can be derived, are solely quoted with statistical 
uncertainties. The TASSO collaboration used for their investigations a 
limited the $x$ range of $0.02$-$0.3$ only.
\footnotetext{Version 7.4 with tuned parameters from 
         \protect\cite{bib-OPAL-ZPC69-543}}

Even 
though the application of the sum rule, Eq.~(\ref{eqn-sumrule}), should 
absorb all the details about the fragmentation functions a large difference 
exists between 
the JETSET expectations at parton and at hadron level. 
This indicates a substantial hadronization correction which is expected to 
behave as a leading order $1/\sqrt{s}$ power correction~\cite{bib-sigma_L-powcorr}. 

For the study of the power correction to the longitudinal cross-section
we used the parametrization of~\cite{bib-NPB511-396} 
where the power correction is given by
\begin{equation}
\label{eqn-powcor-formula}
   \frac{\sigma_L}{\sigma_{\mathrm{tot}}} = 
   \left( \frac{\sigma_L}{\sigma_{\mathrm{tot}}}\right)_{\mathrm{PT}} +
   a_{\sigma_L} \cdot\frac{16{\cal M}}{3\pi^2}  
      \frac{\mu_I}{\sqrt{s}} \cdot\left(\alpha_0(\mu_I)-\alpha_S(\mu)+{\cal O}(\alpha_S^2)\right)
     \ \ \ .
\end{equation}
Here ${\cal M}\approx 1.49$ is the Milan factor, $\mu_I$, usually chosen 
to be $2$~GeV, is the infrared matching scale of the non-perturbative term 
with the perturbative terms. The non-calculable parameter $\alpha_0(\mu_I)$ 
is to be determined from a fit to the data.
The coefficient of the power correction for $\sigma_L$ given in 
~\cite{bib-PLB352-451,bib-JHEP9911-025} is $a_{\sigma_L}=\pi/2$ 
when being adapted for the parametrization chosen in 
Eq.~(\ref{eqn-powcor-formula}). Since the available 
$\sigma_L/\sigma_{\mathrm{tot}}$ data are not sufficiently precise for 
a detailed test of the power corrections, we solely quote for illustrative
purposes the result $\alpha_S(m_{\mathrm{Z}})$ and 
$\alpha_0(2\ {\mathrm{GeV}})$ from fitting the second order plus power 
correction prediction.  This yielded:
\begin{equation}
\label{eqn-powcor-results}
       \alpha_S(m_{\mathrm{Z}})    = 0.126 \pm 0.025 
{\mbox{\ \ \ and \ \ \ }}
       \alpha_0(2\ {\mathrm{GeV}}) = 0.3 \pm 0.3
\ \ \ ,
\end{equation}
where the errors propagated from the total uncertainties of the
$\sigma_L/\sigma_{\mathrm{tot}}$ data. With the large uncertainties
and without further data no definite conclusion about the power 
correction for the longitudinal cross-section is possible. 
The size of the power correction expected at the Z peak was
estimated to be 
$\delta^{\mathrm{pow}}(\sigma_L/\sigma_{\mathrm{tot}}) = 0.010 \pm 0.001$
\cite{bib-PLB352-451} using the measured value of 
$\sigma_L/\sigma_{\mathrm{tot}}$ at $\sqrt{s}\approx m_{\mathrm{Z}}$
quoted in \cite{bib-OPAL-ZPC68-203}.

\section{Summary}
\label{sec-summary}
This paper presents the first measurement of the longitudinal and 
transverse cross-sections at PETRA energies of 35 through 44 GeV. 
Values of
\begin{eqnarray}
      \frac{\sigma_L}{\sigma_{\mathrm{tot}}} & = & 0.067 \pm 0.011 {\mathrm{(stat.)}} 
                                                      \pm 0.007 {\mathrm{(syst.)}}     \\
      \frac{\sigma_T}{\sigma_{\mathrm{tot}}} & = & 0.933 \mp 0.011 {\mathrm{(stat.)}} 
                                                      \mp 0.007 {\mathrm{(syst.)}}   \nonumber
\end{eqnarray}
were obtained
for the longitudinal and transverse cross-sections relative to the total hadronic 
cross-section, where the errors are statistical and systematics.
The two results are exactly anti-correlated since the relative transverse 
cross-section was obtained using the relation~(\ref{eqn-sigma_tot}).
The statistical errors are due to the data and the limited Monte Carlo 
simulation statistics which are about equal in size. 

Using the second order QCD prediction for the relative longitudinal cross-section,
a value of 
\begin{equation}
      \alpha_S(36.6\ {\mathrm{GeV}}) = 0.150 \pm 0.020 {\mathrm{(stat.)}}  
                                             \pm 0.013 {\mathrm{(syst.)}} 
                                             \pm 0.008 {\mathrm{(scal.)}} 
\end{equation}
was determined for the strong coupling 
constant at the luminosity weighted average centre-of-mass energy of $36.6$~GeV. 
Evolved to the Z peak using the 3-loop formula from \cite{bib-PDG} this result 
corresponds to
 $\alpha_S(m_{\mathrm{Z}}) = 0.127 ^{+0.017}_{-0.018}$ 
with total errors, which is in agreement with the current average 
$\alpha_S(m_{\mathrm{Z}}) = 0.1184 \pm 0.0031$~\cite{bib-Bethke-0004021}.

Power corrections to the longitudinal cross-section were considered. The 
available measurements 
at $36.6$~GeV and $91.2$~GeV are not sufficient yet for a definite 
conclusion.

\appendix
\par
\section*{Acknowledgements}
M.B. thanks the members of the III.\ Physikalisches Institut 
of the RWTH Aachen for their generous hospitality and many
helpful discussions.

\end{document}